\documentclass[apj]{emulateapj}

\usepackage{natbib,graphicx}

\usepackage{epsfig}
\usepackage{ulem}
\usepackage{color}
\usepackage{graphics,graphicx}
\usepackage{times} 
\usepackage{amssymb}
\usepackage{amsmath}
\usepackage{url}
\usepackage{multirow}
\usepackage{ctable}
\usepackage{threeparttable}
\newif\ifAMStwofonts
\AMStwofontstrue

\def\aap{A\&A}                   
\def\apjs{ApJS}                  
\def\apjl{ApJ}                   

\def\xmm{{\it XMM-Newton~\/}}

\def\rxte{{\it RXTE~\/}}

\def\suzaku{{\it Suzaku}}

\def\epicmos1{{\it EPIC}{\rm-MOS1~\/}}
\def\epicmos2{{\it EPIC}{\rm-MOS2 ~\/}}
\def\epicmos{{\it EPIC}{\rm-MOS}}
\def\xis{{\rm XIS}}
\def\pin{{\rm PIN}}

\def\xmm{{\it XMM-Newton}}

\def\rxte{{\it RXTE}}

\def\xspec{\hbox{\sc XSPEC}}

\def\heasoftv{\hbox{\rm HEASOFT\thinspace v6.6.1}}
\def\xselect{\hbox{\rm XSELECT}}
\def\ftool{\hbox{\rm FTOOL}}

\def\ks{\hbox{$\rm\thinspace ks$}}

\def\deg{$^{\circ}$}  

\def\pcmsq{\hbox{$\rm\thinspace cm^{-2}$}}

\def\ev{\hbox{$\rm\thinspace eV$}}
\def\kev{\hbox{$\rm\thinspace keV$}}

\def\ctsps{\hbox{$\rm\thinspace count~s^{-1}$}}

\def\ergpcmsqps{\hbox{$\rm\thinspace erg~cm^{-2}~s^{-1}$}}

\def\ergcmps{\hbox{$\rm\thinspace erg~cm~s^{-1}$}}

\def\rg{${\it r}_{\rm g}$}
\def\rin{${\it r}_{\rm in}$}

\def\nh{${\it N}_{\rm H}$}
\def\ka{$K\alpha$}

\def\chisq{{\chi^{2}}}

\def\pl{\rm{\small POWERLAW}}

\def\laor{\rm{\small LAOR}}
\def\gaussian{\rm{\small GAUSSIAN}}

\def\diskbb{\rm{\small DISKBB}}
\def\compps{\rm{\small compPS}}

\def\reflionx{\rm{\small REFLIONX}}
\def\refbhb{\rm{\small REFBHB}}

\def\smedge{\rm{\small SMEDGE~\/}}
\def\kdblur{\rm{\small KDBLUR}}

\def\relconv{\rm{\small RELCONV}}

\def\pexriv{\rm{\small PEXRIV}}

\def\xspec{\hbox{\small XSPEC~\/}}

\def\heasoftv{\hbox{\rm{\small HEASOFT}~v6.11.1\/}}
\def\xselect{\hbox{\rm{\small XSELECT~\/}}}

\def\ftool{\hbox{\rm{\small FTOOL}}}

\def\grppha{\hbox{\rm{\small GRPPHA~\/}}}

\def\mathpha{\hbox{\rm{\small MATHPHA}}}
\def\addascaspec{\hbox{\rm{\small ADDASCASPEC~\/}}}

\def\hxddtcor{\hbox{\rm{\small HXDDTCOR}}}

\def\addascaspec{\hbox{\rm{\small ADDASCASPEC}}}

\def\grid25{\hbox{\rm{\small GRID25}}}
\def\aeattcor{\hbox{\rm{\small AEATTCOR}}}
\def\pile_est{\hbox{\rm{\small PILE_EST}}}
\def\xistime{\hbox{\rm{\small XISTIME}}}
\def\xiscoord{\hbox{\rm{\small XISCOORD}}}

\def\j1118{\hbox{\rm XTE J1118+480}}
\def\j1749{\hbox{\rm J17497-2821}}

\def\je{\hbox{\rm XTE~J1652--453}}

\def\j1752{\hbox{\rm XTE~J1752--223}}

\def\maxi{\hbox{\rm MAXI~J1836--194}}

\begin{document}
\title[Suzaku observation of \maxi] {\suzaku\ Observation of the Black Hole Candidate \maxi\ in a Hard/Intermediate Spectral State} 
\author{
R.~C.~Reis\altaffilmark{1},
J.~M.~Miller\altaffilmark{1},
M.~T.~Reynolds\altaffilmark{1},
A.~C.~Fabian\altaffilmark{2} and
D.~J.~Walton\altaffilmark{2}
}
\altaffiltext{1}{Dept. of Astronomy, University of Michigan, Ann Arbor, Michigan~48109, USA}
\altaffiltext{2}{Institute of Astronomy, University of Cambridge, Madingley Rd., Cambridge CB3 0HA, UK}

\begin{abstract}  We report on a \suzaku\ observation of the {newly} discovered X-ray binary \maxi. The source is found to be in the \textit{hard/intermediate} spectral state and displays a clear and strong relativistically broadened iron emission line. We fit the spectra with a variety of phenomenological, as well as physically motivated disk reflection models,  and find that the breadth and strength of the iron line is always characteristic of emission within a few gravitational radii around a black hole. This result is independent of the continuum used and strongly points toward the central object in \maxi\ being a stellar mass black hole rotating with a spin of  $a=0.88\pm0.03$ (90\% confidence). We discuss this result in the context of spectral state definitions,  physical changes (or lack thereof) in the accretion disk and on the potential importance of the accretion disk corona in state transitions.\\
\end{abstract}


\section{Introduction}
\label{introduction}

The X-ray spectra of X-ray binaries provide important clues on the nature of the compact objects and on the broad properties of the accretion flow. In particular, the various reflection features endemic to stellar mass black hole binaries in all active states  have been successfully used to constrain the dimensionless spin parameter of various black holes in binary systems, as well as providing invaluable insight into the manner in which accretion flow varies with mass accretion rate (e.g. \citealt{miller07review}).  By far the most prominent -- and probably the most important -- of these reflection features is the relativistic iron line appearing at approximately 6.4 to 6.97\kev\ depending to the ionisation state of the emitting material (e.g.~\citealt{tanaka1995}).  \suzaku\ combines exceptional energy resolution below $\approx 10$\kev\ with broadband observation, and, as such, is unique amongst other current X-ray satellites in the study of black hole transients.

The observed spectrum often exhibit the presence of thermal emission, originating in an optically-thick accretion disk together with a hard component often referred to as the corona, and reflection features. There is a complex link between the disk, coronal hard X-rays and reflection emissions (e.g.~\citealt{donereview2007}), which  manifests in various spectral states (see \citealt{Remillard06} and \citealt{Bellonibook2010}). Characterising the driving force between these state transitions is a fundamental challenge for both theoretical and observational studies of accretion-flow properties.
 
The prevailing paradigm requires that in quiescence (very low $\dot{m}$), the inner accretion disk is fully replaced by an Advection Dominated Accretion Flow (ADAF; e.g.~\citealt{Esin1997}). This has led to the idea that the transition between active states is a manifestation of  changes in the innermost extent of the accretion disk, so  that a transition from the disk-dominated \textit{high/soft} state to  a powerlaw-dominated \textit{low/hard} state marks the point of the disk recession. The constant presence of radio jets in the \textit{low/hard} state can also be associated, albeit in a qualitative manner,  with the truncation radius. Thus, it is clear that knowledge of the inner extent of the accretion disk can have fundamental consequences to our understanding of the nature of the  accretion flow at low $\dot{m}$  as well as the connection between accretion disk, corona and radio jets. This radius can be determined via the study of both the continuum emission from the accretion disk or by the reflected iron emission line, again making \suzaku, with its broad band coverage and high spectral resolution,  ideal for this science. In fact, the advent of \xmm\ and \suzaku\ has strongly challenged the paradigm that the accretion disk is  truncated in the bright phases of the \textit{low/hard} state in black hole binaries (but see \citealt{donetrigo2010}; also see \citealt{pileup2010}).

This challenge is exemplified by the recent \suzaku\ and \xmm\ observations of \j1752\ \citep{reis1752}  and the 42\ks\ \xmm\ observation of \je\ \citep{hiemstra1652}.  \j1752\  was caught  during the decay of its 2009 outburst in both the {\it intermediate} (\suzaku) and \textit{low/hard }(\xmm) spectral states.  Interestingly, in both observations we found the presence of a strong, relativistic iron emission line which independently yielded strong constraints on the inner radius: $R_{inner} = 2.8^{+0.3}_{-0.1}$\rg\ and $3.9\pm 0.5$\rg\ (90\% confidence; \citealt{reis1752}) for the \textit{intermediate} and {\it low/hard} state respectively, as well as thermal disk components clearly following the  $L\propto T^4$ relation expected for geometrically-thin accretion disk, thus strongly ruling out disk truncation in either state.  Similar results have been found for \je, where \citet{hiemstra1652} find the disk to be at $\approx 4$\rg\ in the \textit{hard/intermediate state}, and for GX~339-4 where the disk does not appear to truncate until at least $10^{-3}L_{Edd}$ (\citealt{Millergxlhs2006, reisgx, tomsick09gx, wilkinsonuttley09}).   

In order to determine whether these sources are anomalous, or if state transitions are really not linked with a recession of the innermost extent of an optically-thick, geometrically-thin accretion disk, we urgently need observations at low fractions of the Eddington limit, down to $10^{-4}$ and below. However, it is clear that at in some phases of the \textit{ low/hard} state, at least as defined by \citet{mcclintock06}, the disk does not appear to truncate beyond the radius of the Innermost Stable Circular Orbit (ISCO). Whether this is due to a different phase of the \textit{low/hard} state -- i.e. an  ``ADAF state" -- or whether the disk only truly begin to truncate at much lower $\dot{m}$ remains to be seen. Either way, it is clear that the disk plays an important role in state transitions and even jet creation, be it by physically truncating and allowing for the existence of an inner ADAF zone or by dissipating less gravitational energy and allowing for a more powerful accretion disk corona.

The shape of the iron line is determined by the relative depth of the disk within the potential well of the black hole and, as such, conveys information on its spin. Understanding the role of black hole spin ($a=cJ/GM^2$, $-1\leq a \leq 1$) in shaping accretion flows onto and jets from black holes is an important  goal,  having strong repercussion  in all  areas of astronomy. Stellar-mass black holes, for example,  are likely to gain most of their angular momentum during birth, and their spin is a consequence of the supernovae that results in the creation of the central black hole (see e.g. \citealt{millersn11}). Knowing the spin distribution for these objects thus provides a window into the nature of one of the most powerful explosions in the Universe.  At present, we have approximately a dozen spin measurements made by the use of the relativistic iron lines (e.g. \citealt{miller08gx, miller09spin, reisgx, reis09spin, reis1752}), and a handful obtained from the thermal disk continuum (\citealt{mcclintock06, gou09}; \citealt*{ Steiner2011}). However, in order to make any claim on the possible role of spin on, for example, radio jet power \citep{fender2010jets}, we need to increase our spin demographics. 

In this paper, we draw on the recent \suzaku\ TOO observations of the nearly discovered  black hole candidate \maxi\ to learn about the nature of the innermost accretion flow in this source, and to increase our black hole  spin demographics. The following section summarises all the observations of the source and details the current \suzaku\ observation. Section 3 begins by exploring some of the more phenomenological models used to explain the spectra of X-ray binaries and confirms the black hole nature of the central source.  We conclude this section by using a fully self-consistent and physically motivated model to estimate the spin parameter of the black hole in \maxi.  Section 4  summaries our results and discusses the implications for current ideas of black hole state transitions and interpretations.

\section{Observation and Data Reduction}
\label{observation}

MAXI~J1836--194 was discovered by the MAXI/GSC observatory on 2011 August 30 \citep{atel3611}. Its current evolution,  at various wavelengths, have been reported in various ATels \citep{atel3613, atel3614, atel3618, atel3619, atel3626, atel3652, atel3656, atel3689}  with the latest  report by \citet{atel3689}  strongly suggesting that \maxi\ is indeed a black hole X-ray binary based on VLT mid-IR detections. Similar conclusions were made by \citet{atel3628}  based on EVLA radio detections. The 6\ks\ \rxte/PCA observation of \maxi\ reported by \citet{atel3618} had the spectrum described by an absorbed powerlaw with photon index of 1.84. The authors reported the presence of an iron emission line at $6.3\pm0.2$\kev\ together with a smeared edge at $7.2\pm0.2$\kev\ and alluded to a reflection interpretation of these features. The quoted 3--20\kev\ flux of $9.8\times10^{-10}\ergpcmsqps$ is remarkably similar to that of XTE~J1752--223 in a similar state ($\approx11\times10^{-10}\ergpcmsqps$; $\Gamma=1.83\pm0.02$; \citealt{reis1752}).

\begin{figure}
{\hspace*{-0.5cm}
 \rotatebox{0}{
{\includegraphics[width=8.5cm]{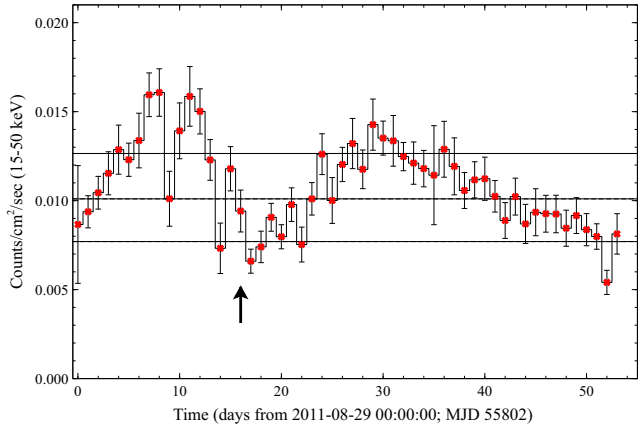}  
}}}
\caption{\label{fig1}Swift/BAT  lightcurve for \maxi\ in the 15--50\kev\ energy range.  The dashed horizontal lines shows the average rate and +/- 1 sigma standard deviation for this source since its discovery. The arrow shows the time of the \suzaku\ observation reported in this paper. Lightcurve available from:  http://heasarc.gsfc.nasa.gov/docs/swift/results/transients/weak/MAXIJ1836-194/}  
\end{figure}

Figure~\ref{fig1} shows the hard X-ray evolution of  \maxi\ as observed by the Swift/BAT Hard X-ray Transient Monitor provided by the Swift/BAT team. \suzaku\ observed the source on 2011 September 14 for  a total of 10.7\ks\ during a period where the 15--50\kev\ flux showed a slight decrease. The three operating detectors constituting the X-ray Imaging Spectrometer (XIS; \citealt{SUZ_XIS}) on-board of \suzaku\ were operated in the 1/4 window ``burst'' mode with both front and back illuminated detectors in the 3x3 and 5x5 editing modes. Using the latest \heasoftv\ software package we processed the unfiltered event files for each CCD following the \suzaku\ Data Reduction  Guide\footnote{http://heasarc.gsfc.nasa.gov/docs/suzaku/analysis/}. {  Due to the observation having been preformed in ``burst'' mode, we started by producing detailed good time intervals (GTIs) using the \ftool\ \xistime\ and setting the option  ``bstgti=yes".} New attitude files were then created using the \aeattcor\  script\footnote{http://space.mit.edu/cxc/software/suzaku/aeatt.html} (\citealt{aeattcor})  in  order to correct for shift in the mean position of the source caused by the wobbling of the optical axis.  The \ftool\ \xiscoord\ was used to create new event files which were then further corrected by re-running the \suzaku\ pipeline with the latest calibration, as well as the associated screening criteria files. The good time intervals provided by the \xis\  team were also employed in all cases to exclude  any possible telemetry saturations.  \xselect\ was used to extract spectral products from these event files. 

In order to estimate the level of pile up suffered by the data we used the   script {{\rm{\small PILE\_EST}}} \footnote{http://space.mit.edu/cxc/software/suzaku/pest.htm}  (\citealt{pile_estimate})  to create a pileup map out of a \suzaku\ event data file. After experimenting with various extraction regions we choose to employ a box annulus region with a {  width} of  240~pixels ($\sim250$'') and a height of 290~pixels ($\sim302$'') and an inner radius of 70~pixels ($\sim73$''). This resulted in a maximum {  pileup fraction of  2~per~cent}. Background spectra were extracted from a circular region having a radius of 100\arcsec\  elsewhere on the same chip.  Individual ancillary response files (arfs) and redistribution matrix files (rmfs) were produced with the script {{\rm{\small XISRESP}}}{\footnote {http://suzaku.gsfc.nasa.gov/docs/suzaku/analysis/xisresp}} -- which calls the tools {{\rm{\small XISRMFGEN}}} and {{\rm{\small XISARFGEN}} -- with the ``medium''  input.

Finally, we combined the spectra and response files from the two front-illuminated instruments (XIS0 and XIS3) using the \ftool\ \addascaspec\ to increase signal-to-noise.  The \ftool\ \grppha\ {  was used to give at least 100 counts per spectral bin in a total of 512 energy channels}.  The nominal energy range covered by  the  \xis\ detectors  is from $\sim0.2-12$\kev. However there are still calibration issues below $\sim1$\kev, which are especially severe in the burst clocking mode. Therefore, we do not consider data below 1.2\kev\ or above 10\kev, following the analyses of \j1752\ presented by \citet{nakahira2011}. The energy band of 1.6--2.4\kev\ is also excluded to avoid large systematics uncertainties in the effective area near the silicon K and gold M edge.  The smaller effective area, together with the fact that the  out-of-time event rate is more significant in the  BI instrument, as compared to the FI, means that at the energies considered here (i.e. above 1.2\kev) the BI data contain even larger uncertainties, and for this reason we did not use XIS1 data in this paper.

We processed the The Hard X-ray Detector (HXD; \citealt{SUZ_HXD})  with the standard criteria. The appropriate response  file (ae{\_}hxd{\_}pinxinome11{\_}20110601.rsp) for XIS-nominal pointing was downloaded{\footnote{http://www.astro.isas.ac.jp/suzaku/analysis/hxd/}}
and the data were reprocessed in accordance with the \suzaku\ Data
Reduction Guide.  As the non-X-ray background (NXB) file was yet to be created by the HXD team at the time of writing, we estimate the NXB by  extracting the earth occulted data'' (ELV$<-5$)\footnote{This has now become available and we have checked the consistency in the results. The "tuned" background has a 15-45\kev\ flux that is approximately 3\% less than the earth-occulted background, and is fully consistency within errors. The spectral shape is also fully consistent with one another and does not alter the results presented here.}. Dead time corrections were
applied with \hxddtcor. The contribution from the Cosmic X-ray Background (CXB)  was simulated using the form of \citet{Boldt87}, with  appropriate normalisation for
the XIS nominal pointing, resulting in a CXB rate of $0.019\ctsps$.
The earth occulted NXB and CXB spectra were then combined using \mathpha\
to give a total background spectrum, to which a 2~per~cent systematic uncertainty was added.  The source spectrum was finally grouped to at least 100 counts per spectral bin. The \pin\ spectrum is restricted to the 15.0--42.0\kev\ energy range and fit simultaneously with the \xis\ data by adding a normalisation factor  {  which is set to 1.16  in all fits with respect to that of the FI spectrum as recommenced by the \suzaku\ data analysis guide. To test the robustness  of our result, in the following sections we also investigate the effect of allowing this cross-normalization to vary}.  All errors reported in this work are 90~per~cent confidence errors obtained by allowing all parameters to vary, unless otherwise noted.

\section{Data Analyses and Results}
\label{simple}
\subsection*{Exploring phenomenological models }

In order to compare the spectral properties of \maxi\ with past work on other X-ray binaries,  we begin by fitting the data with a simple combination of an absorbed powerlaw together with a \diskbb\  \citep{diskbb} model. Figure 2 shows this fit with the 4--7\kev\ range ignored in order to best model the continuum. The total 0.01-100\kev\ unabsorbed flux is $\sim7.8\times10^{-9}\ergpcmsqps$ of which approximately {  42~percent} is associated with the accretion disk. For comparison with the standard work of \citet{mcclintock06} and \citet{Bellonibook2010}, we quote the total 2--20\kev\ unabsorbed flux as $\sim1.5\times10^{-9}\ergpcmsqps$ and a disk fraction of $\approx 0.26$. Combined with a spectral index of  $\sim2.0$,  this observation of \maxi\ is consistent with having caught the source in the \textit{hard/intermediate} spectral state. A further hint of this is also seen in Fig.~1, where it is clear that the source was observed during a time of slight decrease in the hard X-ray flux and suggests a short excursion away from the \textit{low/hard} state.

\begin{figure}

{\hspace*{-0.5cm}
 \rotatebox{0}{
{\includegraphics[width=8.5cm]{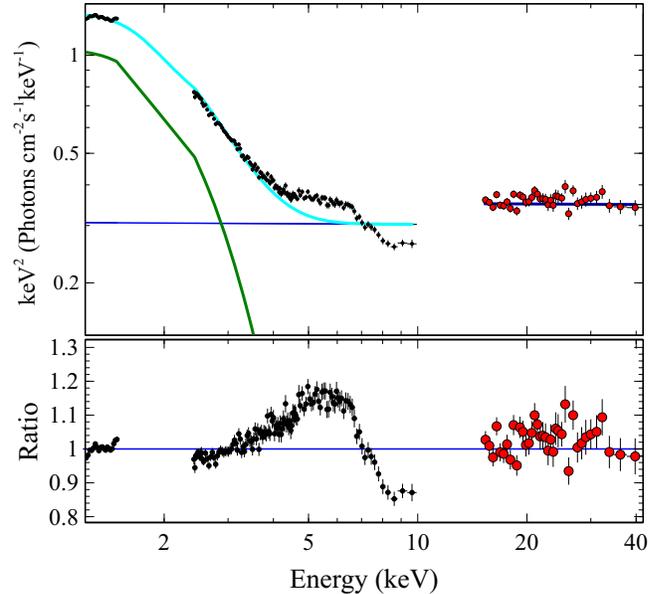}  
}}}
\vspace*{-0.7cm}
\caption{Response unfolded $\nu F\nu$ spectrum of \maxi. The total, disk and powerlaw components are shown as cyan, blue and green solid lines respectively.  {\it Bottom:} Data model ratio to a absorbed \diskbb\  plus \pl\ fit ignoring the 4--7\kev\ energy range.   }
\label{fig2}
\end{figure}

\begin{table*}
\begin{center}
\caption{Phenomenological fits used to determine the robustness of the emission line.}
\label{table}
\begin{tabular}{lcccccccccc}                
  \hline
  \hline 
& {Model~1a} & {Model~1b}& {Model~2}  &{Model~3a}& Model~3b& Model~3c &Model~3d\\

\nh\ (~$\times10^{22}$\pcmsq) & $0.08\pm0.02$ &$0.11\pm0.02$ & $0.05^{+0.01}_{-0.02}$ & $0.05\pm0.03$ &$0.05\pm0.03$ & $0.03^{+0.03}_{-0.02}$& $<0.05$ \\
$E_{smedge}$ (\kev) & 7.11(f) & $5.4^{+0.2}_{-0.1}$ & --- & ---  & ---&---&---\\
$\tau _{smedge}$  & $2.7\pm0.2$ & $3.25\pm0.3$& --- & ---&--- &--- &---  \\
$\Gamma$        & $2.02\pm0.01$ & $2.05\pm0.01$ & $1.890^{+0.003}_{-0.016}$ & $2.13\pm0.06$& $2.17\pm0.03$&  $2.18\pm0.02$&$2.09\pm0.03$\\
$kT_{disk}$ (\kev)  & $0.445\pm0.005$ & $0.433\pm0.005$& $0.474\pm0.003$ &$0.314^{+0.003}_{-0.006}$&  $0.30\pm0.01$& $0.30\pm0.01$& $0.30\pm0.01$ \\
$kT_{electron}$ (\kev)  & --- & ---  & ---&$31^{+3}_{-5}$& $10^{+2}$& $10^{+2}$& $10^{+6}$\\
$\tau_{CompPs}$  & --- & ---  & ---&$0.358^{+0.04}_{-0.03}$&  $0.39\pm0.02$& $0.9^{+0.4}_{-0.1}$ &$0.92^{+0.05}_{-0.18}$\\
$N_{hard}$ & $0.36\pm0.01$ & $0.38\pm0.01$ & $0.219\pm0.001$ & $0.27^{+0.06}_{-0.05}$& $0.33^{+0.04}_{-0.03}$ & $0.33\pm0.03$&$0.29\pm0.03$\\
$N_{diskbb}$ ($\times 10^{3}$)& $4.4\pm0.3$ &$5.1\pm0.4$& $3.54^{+0.03}_{-0.16}$& ---&---&---&---\\
$N_{CompPs}$($\times 10^{3}$) &  --- &---&---&$30^{+4}_{-3}$& $61^{+16}_{-14}$ & $50^{+13}_{-8}$&$53^{+10}_{-4}$\\
$E_{gauss}$ (\kev) & --- & $6.4-6.97$& --- & ---& $5.5\pm0.1$ & ---&---\\
$\sigma_{gauss}$ (\ev) & --- & $620^{+80}_{-100} $& --- & ---  & $800^{+160}_{-90}$ & ---&---\\
$E_{laor}$ (\kev) & --- & ---& $6.97_{-0.04}$ &---&  ---&  $6.58^{+0.14}_{-0.09}$&$ 6.48^{+0.10}$\\
$N_{line} $  ($\times 10^{-3}$ ) &  --- & $2.6\pm0.7$ & $15.4^{+0.4}_{-0.3}$ &---& $2.8^{+0.8}_{-0.5}$ & $3.2^{+0.3}_{-0.4}$&$1.5^{+0.4}_{-0.3}$\\
$W_{line}$ (\ev) & --- & $300\pm90$ & $360\pm10$ & ---  & $270^{+80}_{-40}$ &$270^{+20}_{-40}$ &$180^{+60}_{-40}$\\
$q_{in}$  & --- & --- & $6.8\pm0.2 $ & ---& ---&$3.8^{+0.3}_{-0.2}$& $3.2\pm0.2$\\
$\theta$ (degrees)  & --- & ---  & $44\pm1$& --- &   ---&25(f)& 25(f)\\
\rin\ (\rg) &--- & ---  &$<1.5$  &  ---& ---&$3.05^{+0.26}_{-0.06}$ & $3.3^{+0.7}_{-0.5}$\\
$\xi$  & --- & --- & --- & $<120$ & $<100$ & $70^{+80}_{-60}$& $5200^{+2600}_{-2200}$\\
$\chi^{2}/\nu$ & 630.9/328 & 380.3/324 & 432.1/324 & 572.5/326 & 372.3/323  &356.8/322&  360.0/322\\ 
  \hline 
\end{tabular}
\end{center} 
\small Notes: Results of phenomenological fits with a variety of  continuum models. The continuum in Model~1 is assumed to consist of a simple powerlaw having a normalisation $N_{hard}$ and a \diskbb\  component. The feature around the 6--7\kev\ range is modelled with a smeared edge (Model~1a) together with a further Gaussian line (Model~1b). Model 2 replaces the smeared edge and the Gaussian line with a single relativistic line component. Model~3 replaces the \diskbb  and powerlaw models with the Comptonization code of \citet{compps} and the model \pexriv\ of \citet{pexrav} respectively. 
For Model~3, a ``slab" geometry was assumed and the reflection option in \compps was deactivated. A lower limit for the electron temperature of 10\kev\ was imposed. The feature is modelled with a broad Gaussian (Model~3b) and a relativistic line (Model~3c). Model~3d self consistently convolves the \pexriv\ model with the same parameters as the relativistic line since they both originate from the same region.    All errors are 90~per~cent confidence.
\end{table*}

In Figure~2, the 4--7\kev\ energy range is shown above the continuum in order to  highlight the presence of various features in this range. A possible explanation used by a number of authors to account for the residuals seen in Fig.~2 assumes that these features are a combination of a narrow iron \ka\ emission line and its associated absorption edge, which, when arising from the region around a black hole, suffers from a high degree of smearing and thus is better described by the \smedge\ model (e.g.  \citealt{smedge}).  A fit with such a smeared edge having an energy of 7.11\kev\ (as expected from neutral iron) and a width of 10\kev\   was not able to account for the residuals ($\chisq/\nu=630.9/328$ Model~1a; Table 1 and Figure 3), with broad residuals remaining both above and below the edge energy. Adding a {  narrow} ($\sigma = 1\ev$)  \gaussian\ line at 6.4\kev\  {  did not improve the residuals in any way}. A  much better fit is indeed achieved when the energies of both the  \gaussian\ line and that of the smeared edge are allowed to be free {\it and} {  the emission line is allowed to be broad} ($\chisq/\nu=380.3/324$; Model~1b; Table 1). The neutral hydrogen density found in these models are also mildly consistent with the value of $2.0\pm0.4\times10^{21}\pcmsq$ presented by \citet{atel3613} based on a Swift/XRT observation. We will discuss possible reasons for the variation in \nh, as  observed between the different models, in \S~4. The scenario so far presented here however, has an \textbf{edge energy of $\sim5.4$\kev} which is much less than the value for neutral iron absorption; itself a lower limit since iron is likely to be highly ionised in the inner parts of the accretion disk. In fact, it is more likely that the edge is trying to compensate for a broad emission feature (see \S~4) and,  for this reason, although the model is detailed in Table~1, we do not believe it to convey any physical information.

Given the obvious presence of a broad line feature in Figure~2, we replaced the \smedge model with a \laor\ line profile \citep{laor} as expected if emission is coming from the inner disk around a black hole.  The line energy is constrained to lie between 6.4 and 6.97\kev, thus encompassing the full range of possible ionisation states of iron. We start in Model 2 with a  {simple powerlaw emissivity} profile such that $\epsilon(r)\propto r^{-q_{in}}$.  The outer disk radius was frozen at the maximum value in the model of 400\rg. This model is detailed in Table~1 and resulted in a satisfactory fit with ($\chisq/\nu=432.1/324$). Allowing for a broken powerlaw emissivity profile with indices $q_{in}$ within a radius $r_{break}$ and $q_{out}$ beyond, further improved the fit ($\Delta \chi^2 = 13.5$ for 2 degrees of freedom).  {This fit has an emissivity index of  $q_{in} \sim7$ within a radius of $\sim6$\rg\  and then breaks to $q_{out}\sim 2.6$. In both these instances, the inner radius obtained is very low implying that not only is the central object in \maxi\ a stellar mass black hole, it is also likely to be rapidly spinning.}  In all models considered so far, the \diskbb\ component consistently required a disk with a temperature of approximately 0.45\kev.

\begin{figure}

{\hspace*{-0.5cm}
 \rotatebox{0}{
{\includegraphics[width=8.8cm]{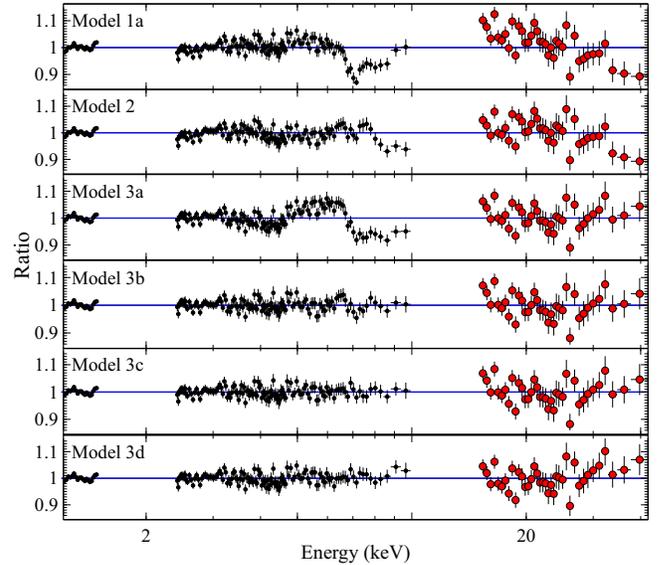}  
}}}
\caption{Data/Model ratio for the various phenomenological models described in Table~1. {\it From top to bottom:}  $phabs* smedge* (diskbb + powerlaw)$  with the smeared edge energy frozen at 7.11\kev; $phabs*(laor + diskbb + powerlaw)$; $phabs*(compps + pexriv)$; $phabs*(compps + pexriv + gaussian )$; $phabs*(compps + pexriv +  laor)$; $phabs*(compps + kdblur\otimes pexriv +  laor)$.  }
\label{fig3}
\end{figure}

\begin{figure*}
\hspace*{-0.5cm}
\rotatebox{270}{\includegraphics[width=4.5cm]{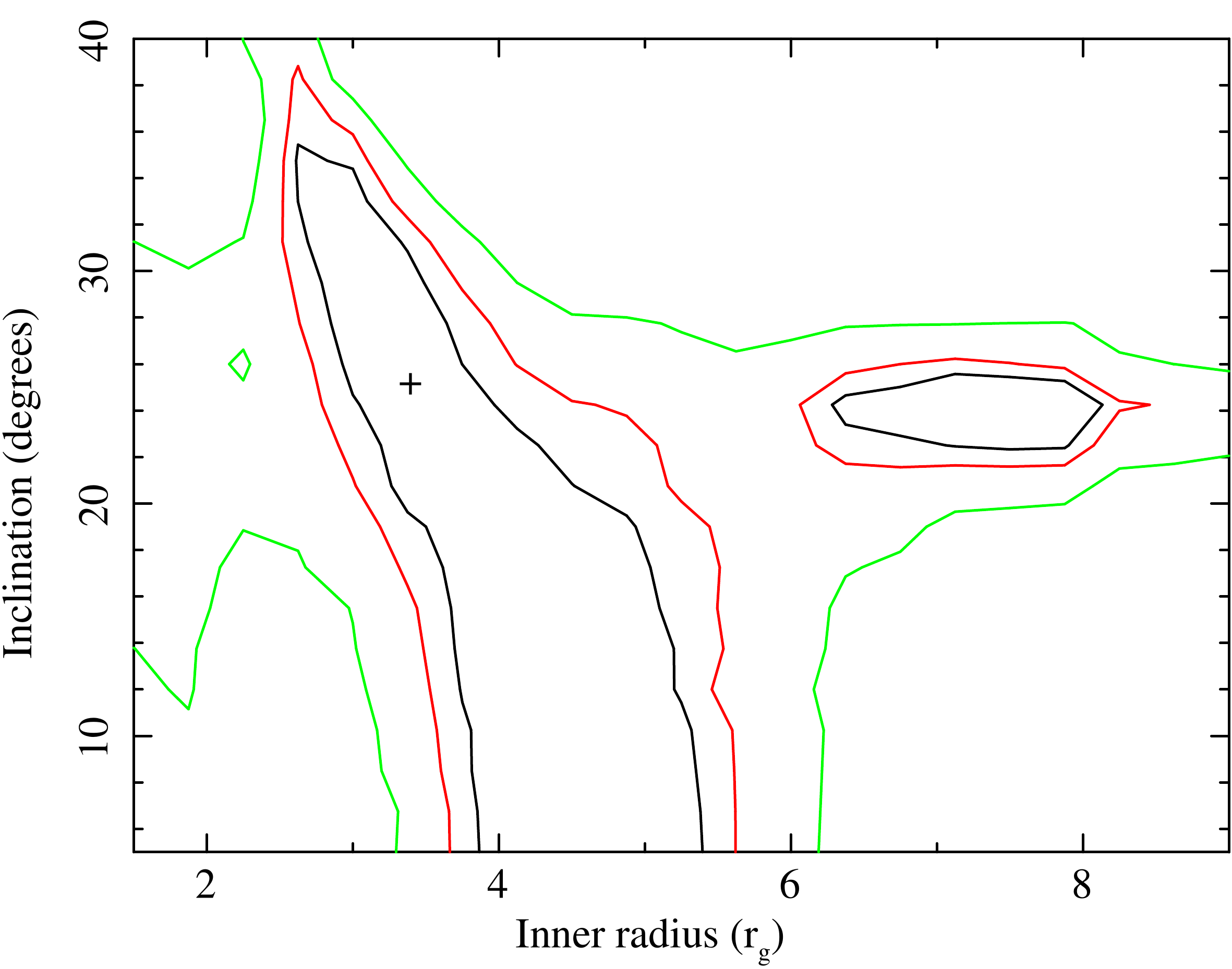}}
\hspace*{0.2cm}
\rotatebox{270}{\includegraphics[width=4.5cm]{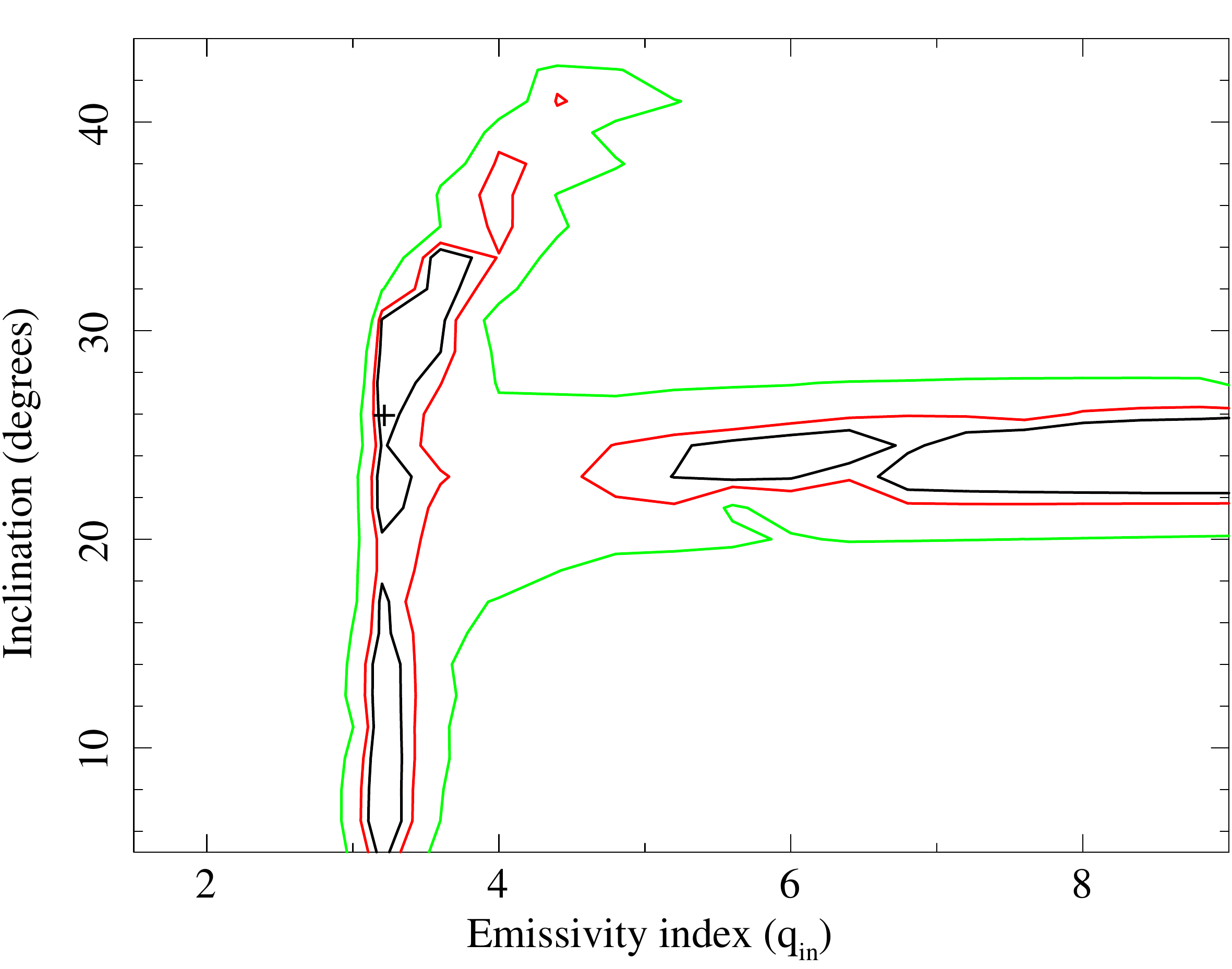}}
\hspace*{0.2cm}
\rotatebox{270}{\includegraphics[width=4.5cm]{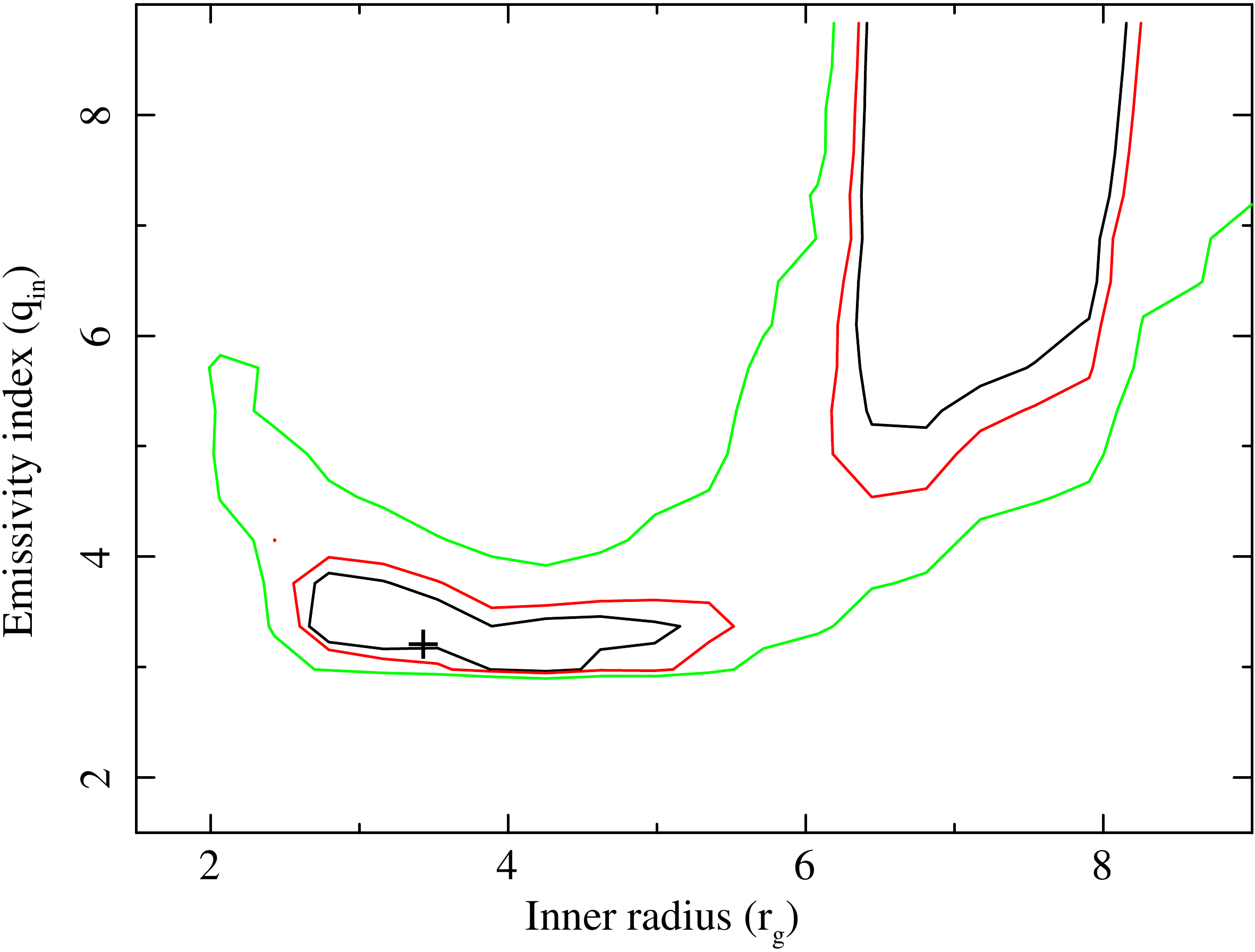}}

\caption{Contour plots investigating the effect of the inclination (left) and emissivity index (right) on the inner radius for Model~3d. The centre panel shows the contour plot for the inclination versus emissivity index. The 68, 90 and 95 per cent confidence range for two parameters of interest are shown in black, red and green respectively.  The cross marks the global minima. It is clear that even for the most curved continuum where the emission feature would appear the narrowest (i.e. Model~3), the inner extent of the accretion disk, as obtained solely from the phenomenological fit to the iron line profile,  is still consistent with not being truncated far beyond the innermost stable circular orbit for a Schwarzschild black hole (6\rg) in the \textit{hard/intermediate} spectral state observed here. However, due to the local minima in these phenomenological models we cannot rule out a disk that is mildly truncated at this stage. }
\label{fig4}
\end{figure*}

It has been suggested by a number of authors that this simple view of a multicolour disk is not appropriate and should be replaced with a broader disk model. \citet{Kolehmainen2011} further suggested that the broad iron line constantly seen in a number of stellar mass black hole binaries might be an artificial effect  caused by the usage of narrower disk components similar to {\small DISKBB}, used here.  Despite the noticeable problems with this interpretation --- iron lines are mostly seen in black hole binaries in the \textit{low/hard} state where the disk temperature is relatively cold and its contribution to the spectrum at $\approx6$\kev\ is unimportant --- we nonetheless test the effect this may have  on the residuals seen between 4--7\kev\ (Figure 2), by  replacing both the \diskbb, \laor,  and \pl\  components with \compps\ \citep{compps} and \pexriv\ \citep{pexrav}. The former fully characterises the process of Comptonization for a variety of coronal geometries, electron distributions and seed photons injection geometries. We used the code assuming a  purely thermal electron distribution in the corona (Gmin=-1).  The \pexriv\ model represents a exponentially cut-off power law  spectrum\footnote{The cut-off energy is frozen at 300\kev, similar to the value used in the reflection model \reflionx\ described in detail in the next section.}} reflected from ionized material. When the reflection fraction, $R>0$, this models give the sum of the illuminating powerlaw hitting the ionised disk together with the corresponding reflection component (Fe-edge and Compton hump) but it \emph{does not} include the Fe~\ka\ emission line. {  We initially have the ionisation parameter ($\xi = 4\pi F/n ~\ergcmps,$ where $F$ is the incident flux and $n$ is the number density of hydrogen nuclei),} reflection fraction and inclination of the \pexriv\ component frozen at $1000\ergcmps$, 1 and 25~degrees respectively, in line with the values used for \j1752\ by \citet{nakahira2011}.  The geometry of \compps\ is chosen to be a ``slab" (geom=1) and the covering fraction frozen at 1. Reflection from the \compps\ component is turned off (rel\_refl=0).

This combination  resulted in a unsatisfactory fit with $\chisq/\nu=629.4/327$. Allowing the ionization of \pexriv\ to vary did not resolve the problem ($\chisq/\nu=572.5/326$),  with a broad feature clearly  present in the residuals between  4--7\kev\ (Model 3a in Table 1; Figure 2). {  Further allowing the reflection fraction to change again did not provide a satisfactory fit  ($\chisq/\nu=546.8/325$). The reflection fraction artificially shoots up to $\sim2.5$ as the} \pexriv\ {  model tries to compensate for the lack of an Fe-emission line  by increasing the depth of the iron-edge (see Figure~9)}. Indeed, adding a broad \gaussian\ line  (Model~3b) resulted in a dramatic improvement, with $\Delta \chisq = 200.2$ for 3 degrees of freedom compared to Model~3a. However, a line energy of $E_{Gauss} = 5.5\pm0.1$\kev\  is not consistent with emission of iron. Furthermore,  a width of  $\sim 800$\ev\ is  highly suggestive of emission from close to the black hole where gravitational broadening effects are important, again taking us back to the need of a relativistic iron line.

We therefore proceed by replacing the \gaussian\ line in Model~3b with the relativistic line expected around a spinning black hole. The inclination in the \laor\ model is  frozen at 25~degrees, as with \pexriv. This model immediately improved the quality of the fit ($\Delta \chi^2 = 15.5$ for 1 degree of freedom) and, more importantly, brought the emission line energy to a range consistent with emission from iron ($E_{Laor}\approx6.6$\kev). The inner radius as obtained from the iron line profile in Model~3c is again consistent with the central object in \maxi\ being a rotating black hole. {  Allowing the reflection fraction in the }\pexriv\ {  model to be free yields $R=0.9\pm0.2$,  and does not change any of the other fit parameters nor does it affect the quality of the fit ($\chisq/\nu=356.4/321$).}  Importantly, Model~3c as it stand is not physically consistent.  If the Fe line is being emitted in the inner parts of the accretion disk, as appears to be the case, then the other reflection features should also experience the effect of strong gravity. For this reason  we convolve the \pexriv\ model --- which models the illuminating continuum  as well as the absorption edge of iron and the  Compton reflection hump --- with the  relativistic kernel \kdblur\ \citep{laor}. We force the parameters of \kdblur\ to be the same as that of the relativistic iron line. {  This model (Model~3d) result in a slightly worst quality of fit $\Delta \chisq = 3.2$ for the same number of degrees of freedom as the previous model, however it is now physically consistent.  Again, allowing the reflection fraction in the} \pexriv\ {  model to be free resulted in an improvement of $\Delta\chisq = 4.4$ for 1 degree of freedom, giving  $R<0.91$ but did not change the line profile in any way, with  the inner radius and emissivity index remaining at \rin$=3.3^{+0.7}_{-0.4}$\rg\  and  $q_{in}=3.4\pm0.2$, respectively.  We also investigated the robustness of the fit with respect with the cross-normalization constant between the \xis\ and \pin\ data which we have so far fixed at 1.16. Allowing this to go free barely improved the quality of the fit ($\Delta\chisq = 1.4$ for 1 degree of freedom) and  recovered a value of $1.14^{+0.03}_{-0.08}$, fully consistent with the expected value.  }

\begin{table*}
\begin{center}
\caption{ Fits with physically motivated reflection models }
\label{table2}
\begin{tabular}{lcccccccc}                
  \hline
  \hline 
  
& Model~4a & Model~4b & Model~5a & Model~5b  \\

\nh\ (~$\times10^{22}$\pcmsq) & $0.11\pm0.02$ & 0.11(f) & $0.191^{+0.003}_{-0.030}$ &$0.19\pm 0.03$ \\
$\Gamma$ 									&$1.92\pm0.04$  & 1.92(f) &$1.82^{+0.06}_{-0.01}$ & $1.81^{+0.05}_{-0.03}$ \\
$kT_{disk}$(\kev)  						& $0.448^{+0.006}_{-0.008}$ & 0.448(f) & $0.27^{+0.02}_{-0.01}$ & $0.26^{+0.02}_{-0.03}$ \\
$N_{hard}$ 									& $<0.01$   &$ < 0.01$& $0.041^{+0.014}_{-0.002}$ &$0.030^{+0.077}_{-0.004}$ \\
$N_{diskbb}$ ($\times10^3$)		& $4.5^{+0.5}_{-0.4}$&4.5(f)& ---&--- \\
$F_{Illum}/F_{BB}$ 					    & ---& ---&$1.2\pm0.2$& $1.2^{+0.2}_{-0.3}$\\
$n_{H}$ ($\times10^{19}$)      &--- & --- &$2.5^{+0.7}_{-1.0}$ & $2.9^{+1.7}_{-0.6}$ \\
$N_{refbhb}$                               &---  &--- &$0.31^{+0.03}_{-0.04}$  &$0.30^{+0.02}_{-0.04} $\\
$N_{reflionx}$ ($\times10^{-6}$) &$3.7\pm0.4$  &$4.5^{+0.5}_{-0.6} $ &---  &---\\
$q_{in}$ 										&$>5.8$ & $>4.4$ &$>6.5$ & $>7.3$\\
$q_{out}$ 										&$3.4\pm0.1$ & $3.7\pm0.1$ &$3.12^{+0.06}_{-0.12}$ & $3.19^{+0.07}_{-0.05}$\\
$r_{break}$ (\rg)							&$3.6^{+2.3}_{-0.2}$ & $3.2^{+2.1}_{-0.1}$ &$3.4^{+0.2}_{-0.1}$ & $3.6^{+0.2}_{-0.1}$\\
$\theta$ (degrees) 						& $<15$ & $<30$ & $<23$ & $<17$ \\
$\xi$											& $2150^{+170}_{-250}$ & $2170^{+500}_{-250}$ &--- &--- \\
\rin\ (\rg) 									&$2.54^{+0.06}_{-0.20}$ & $1.8^{+0.8}_{-0.3}$&$2.2^{+0.2}_{-0.4}$&$(2.45^{+0.18}_{-0.20})^a$ \\
Spin ($a$) 									&---& --- &--- & $0.88\pm0.03$\\
$\chi^{2}/\nu$                          &379.7/322   & 198.5/211 & 344.9/322 & 344.5/322 \\
\hline
\hline
\end{tabular}
\end{center} 
\small Notes:  Model~4 is described in \xspec\ as $phabs*(diskbb + powerlaw +kdblur*reflionx)$. Model~4b is identical to 4a but we only fit the 3--10\kev\ energy range. Model~5 replaces \reflionx\ and \diskbb\ with the fully self-consistent reflection model \refbhb. In all models described so far the kernel from the \laor\ line profile was used to account for the gravitational effects close to the black hole. In Model~5b, we finally replaces the \kdblur\ kernel with the  relativistic code \relconv\ were the spin is a parameter of the model. In all cases the hard emission illuminating the disk is assumed to be a powerlaw with index $\Gamma$. All errors are 90~per~cent confidence for one parameter. $^a$ Inner radius is not a model parameter and was derived by using the relationship between the spin and ISCO \citep{Bardeenetal1972}. It is shown here merely to allow for easy comparison.
\end{table*}

To expedite the computational time, in all incarnations of  model~3 we have assumed that the inner disk inclination has a value of 25~degrees.  As a last step in our exploration of these phenomenological models,  {we allow the inclination to be free and investigate any possible degeneracy this might have on the inner accretion radius and emissivity index. Figure~\ref{fig4} shows the 68, 90 and 95 percent confidence range for both parameters as a function of inclination as well as the inclination as a function of emissivity index. From Fig.~4, it can be seen that despite the fact that the inclination is not very well constrained, ranging from anywhere between 5 and 40 degrees at the 90 per cent confidence level, the inner radius of the accretion disk, as obtained solely from the breadth of the line profile using purely phenomenological models, is still consistent with being at  or close to the radius of the Innermost Stable Circular Orbit around a rotating black hole. The  global minimum which is marked with a black cross in all panels, still requires $r_{in} < 5.6$\rg\ at the  90~percent level however we do see the presence of a further solution having a radius which is marginally consistent with the ISCO of a non-rotating black hole together with a very high emissivity index.  In Fabian et al. (2012), we showed that  a single power-law emissivity profile has only limited validity and that a slope of $\sim3$ is a fair approximation for an inner disk starting at \rin$\sim2$\rg. However, it severely underestimates
the profile within \rin$\sim2$\rg\ and is therefore a poor probe of the innermost region around a rapidly spinning black hole. In that paper, we argued that if q = 3 is used, then it will likely yield an upper limit to the inner radius and thus a lower limit on the spin if the source is indeed spinning rapidly. The evidence so far points toward the black hole in \maxi\ \emph{not}  being rapidly spinning, and therefore, from theoretical arguments we should expect an emissivity index close to 3, similar to the value found in the the global minima having an inner radius within $r_{in} < 5.6$\rg. In the following section, we will investigate these possible degeneracies fully using a number of self-consistent reflection models.}

In this section, we have established beyond any reasonable doubt the presence of a strong and mildly broad emission line associated with iron in the \suzaku\ spectra of \maxi. We have shown that different continuum models, as well as phenomenological features (such as the blurred edge component)  does not eliminate the need for a broad emission line. Our efforts to ``remove" the need for a broad emission line  are somewhat artificial since reflection is a natural consequence of a system where hard X-rays are impinging on a cold accretion disk. However, it is only by doing so that we can safely rule out any continuum  effect on the breadth of the iron emission line. In the following section, we endeavour to interpret the spectra in a fully consistent and physical manner.

\subsection*{Physically self-consistent modelling}
\label{physical}

In all our previous fits, a broad feature has been shown to be robustly present above a thermal-disk and powerlaw-like continuum, peaking around the energies expected for neutral ${\rm Fe~I~K_\alpha}$ emission ($\sim 6.4$\kev) and highly ionised H-like ${\rm Fe~XXVI~ L}_{\rm y\alpha}$ ($\sim 6.97$\kev; See Fig.~\ref{fig2}). The natural explanation for this feature is that it is indeed associated with the reprocessing of hard X-ray emission by an accretion disk and, as such, the broad feature observed is the signature of iron fluorescence that has been relativistic broadened due to the strong gravity around the central black hole. Similar reflection features are observed in a wide range of objects ranging from neutron stars (\citealt{ bhattacharyya07, cackett08, cackett09, cackett10, disalvo09, reisns}), stellar mass black holes (\citealt{miller07review, blum09, reis09spin, hiemstra1652, Waltonreis2012}) and AGNs (\citealt{ tanaka1995,  FabZog09, miniutti09spin, schmoll09, waltonreis2010, Nardini2011, 3783p1, 3783p2}).

\begin{figure}

{\hspace*{-0.5cm}
 \rotatebox{0}{
{\includegraphics[width=8.5cm]{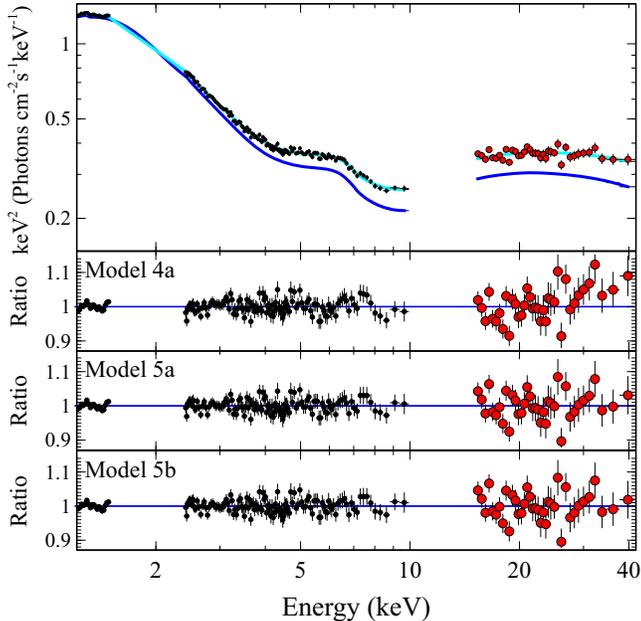}  
}}}
\vspace*{-0.3cm}
\caption{{\it From top to bottom:} Response unfolded $\nu F\nu$ spectrum to the best-fit Model~5b. The total and blurred \refbhb\ components are shown as cyan and blue solid lines respectively. The powerlaw continuum falls below the y-scale;  Data/model ratio for: (4a) $phabs*  (diskbb + powerlaw + kdblur\otimes reflionx)$; (5a) $phabs*(powerlaw + kdblur\otimes refbhb )$ and (5b) $phabs*(powerlaw  + relconv \otimes refbhb)$.}
\label{fig5}
\end{figure}

Due to the immense diagnostic potential  of reflection features, a large theoretical effort has been devoted to fully characterising the reflection spectrum expected to arise from such systems \citep{LightmanWhite1988, George91, Matt1991,  rossfabian1993, Zycki1994, nayakshin2000, cdid2001, reflionx,  refbhb, GarciaKallman2010, GarciaKallman2011}. Amongst these, the most widely  used  reflection model is the \reflionx\ code of \citet{reflionx}. This model self-consistently calculates the reflection arising from all energetically important ionization states and transitions expected in disks around black holes. At low ionization parameters  it reproduces the reflection continuum first described by \citet{LightmanWhite1988}, as well as self-consistently calculating the fluorescent lines; at higher $\xi$, lower Z element becomes ionized which results in a  softening of the reflection spectrum. We start by using a combination of \reflionx, relativistically  convolved with  \kdblur, together with disk emission ({\rm{\small DISKBB}}) and a powerlaw illuminating continuum. The photon indices of the \reflionx\ and powerlaw components are assumed to be the same. The outer disk radius is assumed to be at 400\rg\ (the maximum allowed by the model) and the iron abundance of \maxi\ is  fixed at solar. This model give a poor fit with  $\chisq/\nu=400.6/324$. Allowing instead for a broken powerlaw emissivity profile results in a significant improvement with $\chisq/\nu=379.7/322$. This models is described in  detail in Table~2 (Model~4a) and shown in  Fig.~\ref{fig5}. 

It is worth noting some similarities {  and differences} between the values obtained from the current reflection model and those from the phenomenological fits described in the previous section. To start, the high ionization parameter  found here ($\xi \approx 2200\ergcmps$) is  similar to that found in the most physically motivated version of Model~3  (i.e. Model~3d) and suggest an intermediate to high ionization state. The moderate to low disk inclination shown in Fig.~\ref{fig4}  is also confirmed here, where the current reflection fit suggests $\theta \lesssim 15$\deg. The disk parameters found here are also in close agreement with those found in Model~2, however we cannot compare the parameters with Model~3 as the disk was modelled assuming a Comptonization model. {   A major difference between the fit with the reflection  over that of the various phenomenological models is in the  necessity for a broken emissivity profile over that of a single powerlaw. In model~4a, the disk extends to within $\sim2.5$\rg. The steep emissivity occurs in a very narrow annuli between this radius and $\sim4$\rg, at which point it goes back to the value expected from a purely Newtonian geometry.   It is interesting to note that in the phenomenological model~3d, the inner radius obtained with an emissivity index of $\sim3.3$ is similar to the break radius found here. In the case of the phenomenological models, the inner radius was obtained from a pure \emph{emission line}. The approach of using a model such as the} \laor\ {  line profile to obtain spin has many limitations as the information imprinted  by the effects of strong gravity is \emph{not} limited to an emission line. In fact, such a limitation is clear in the inconsistency between the high ionization found here ($\xi\sim2200\ergcmps$) as well as that found in Model~3d  ($\xi\sim5000\ergcmps$) with the line centroid of $\sim6.48$\kev\ found in that same model which indicates lowly ionised iron. In Model~3d, a further constraint on the inner radius was in place by the act of convolving the} \pexriv\ {  model with the same kernel as the} \laor\ {  line profile. Here, there was an interplay between broadening of  the emission line component as well as the iron absorption edge (modelled separately and therefore not forced to be physical). The final product was the apparent presence of an emission  line at  $\sim6.48$\kev, as expected from cold matter with $\xi \lesssim100\ergcmps$, together with the absorption edge from hot hydrogenic iron Ka FeXXV and/or FeXXVI  with edge energies $\gtrsim8.85$\kev\  as expected from $500 \lesssim \xi \lesssim5000\ergcmps$. This combination acted to model a broad feature with only a mild degree of relativistic broadening.} \reflionx\ {  on the other hand, by virtue of  the imposed self-consistency in its atomic physics, modelled the same broad feature by allowing a greater degree of broadening to be attributed to gravity, and hence resulted in a smaller inner radius.}

\begin{figure}
{\hspace*{-0.5cm}
 \rotatebox{0}{
{\includegraphics[width=8.5cm]{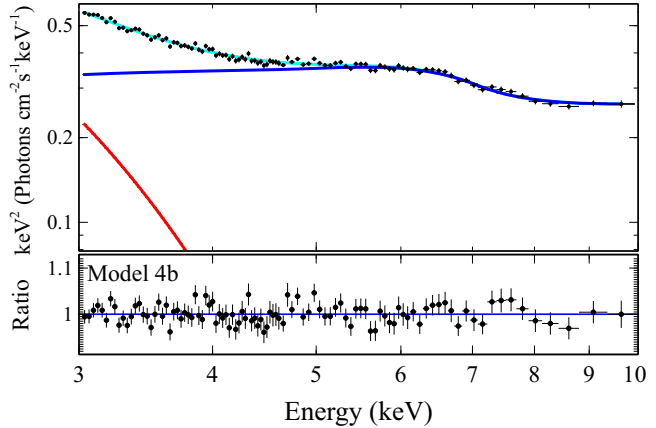}  
}}}
\caption{Response unfolded $\nu F\nu$ spectrum from Model~4b fit in the 3--10\kev\ energy range. The total, blurred reflection and disk components are shown as cyan, blue and red solid lines respectively. The powerlaw continuum falls below the y-scale.  {\it Bottom:} Data model ratio to Model~4b. The reflection parameters obtained in this range are identical to that obtained by modelling the full spectra (see Table~2).}
\label{fig6}
\end{figure}

\begin{figure*}
\hspace*{-0.5cm}
\rotatebox{270}{\includegraphics[width=4.5cm]{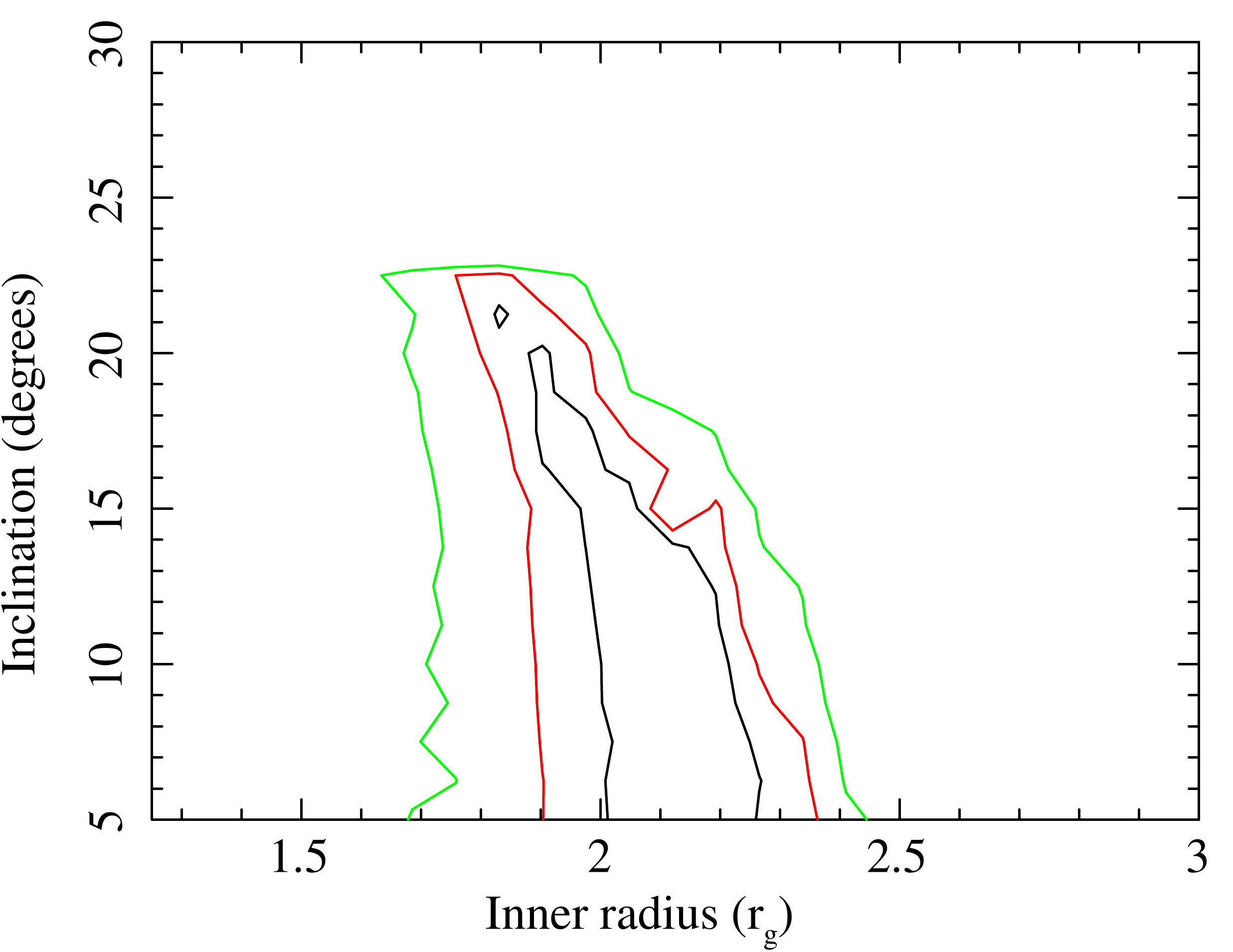}}
\hspace*{0.2cm}
\rotatebox{270}{\includegraphics[width=4.5cm]{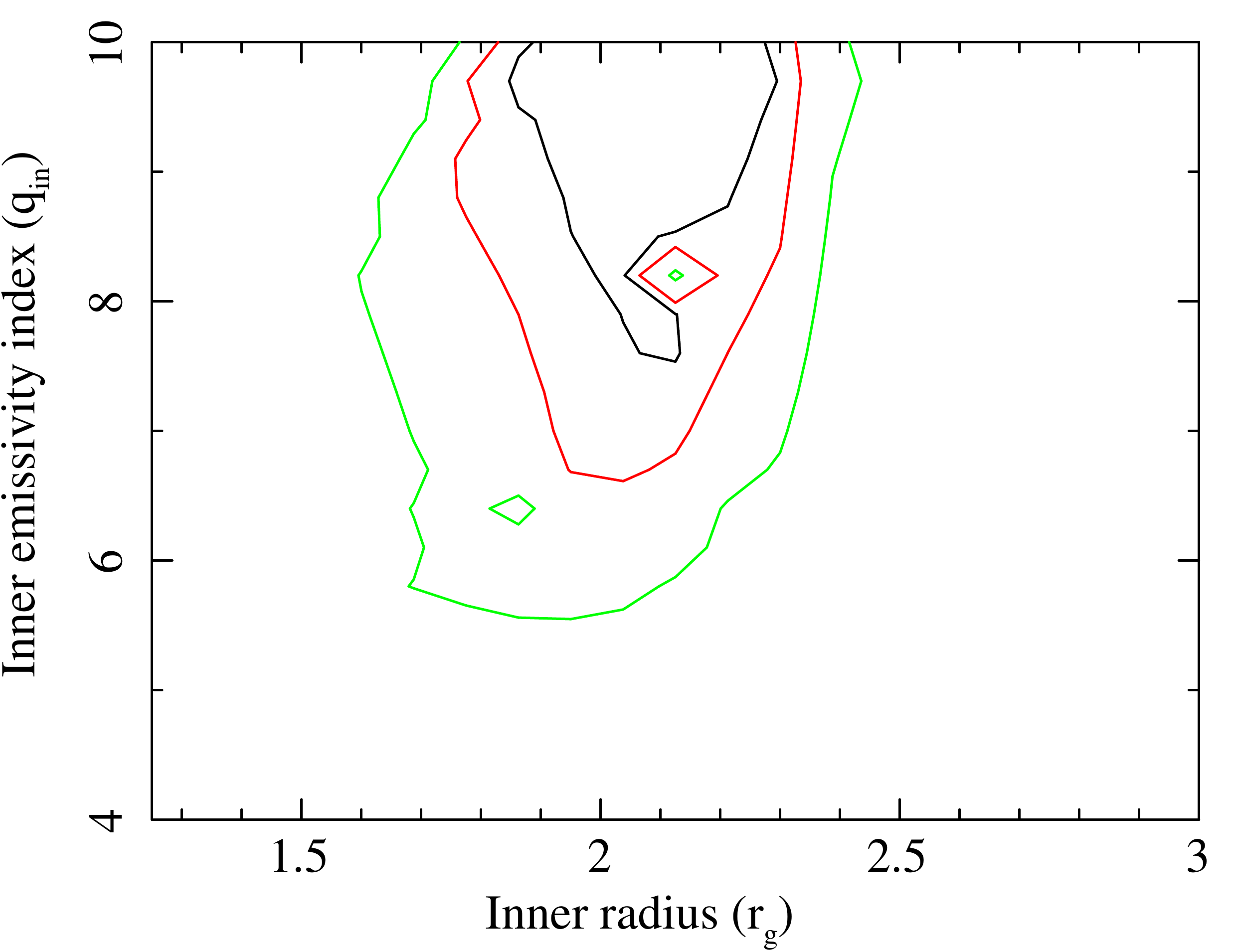}}
\hspace*{0.2cm}
\rotatebox{270}{\includegraphics[width=4.5cm]{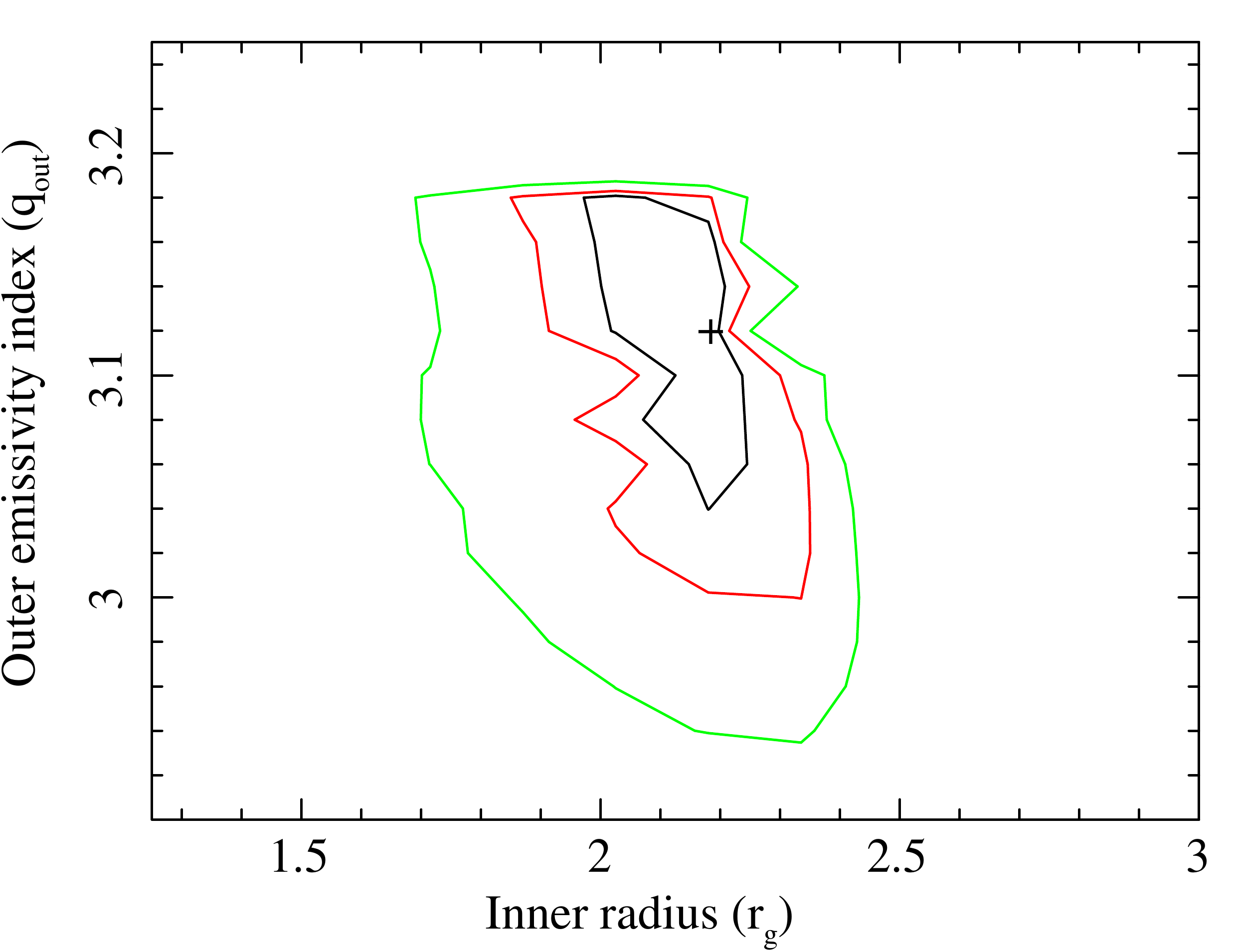}}
\hspace*{-0.5cm}
\rotatebox{270}{\includegraphics[width=4.5cm]{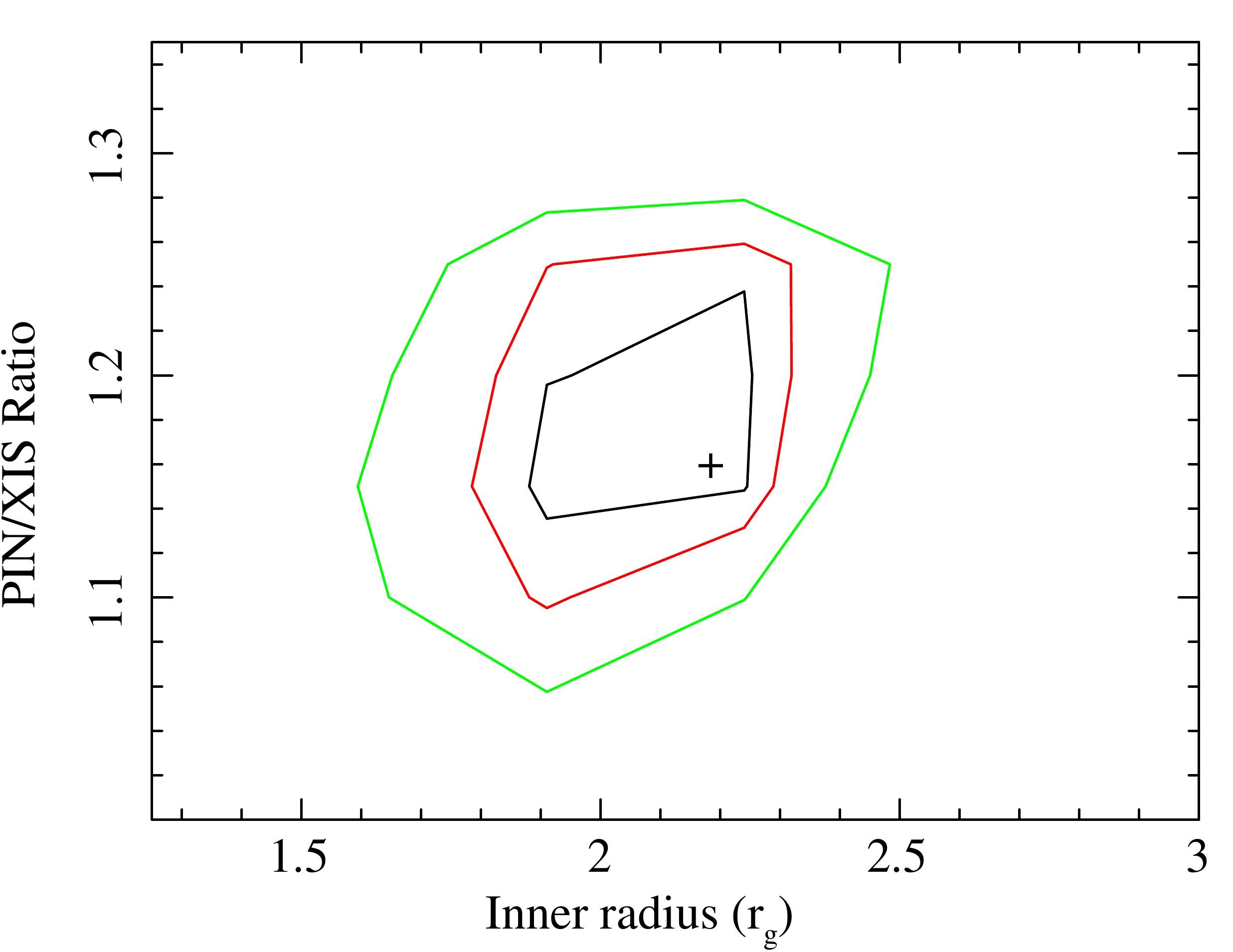}}
\hspace*{0.35cm}
\rotatebox{270}{\includegraphics[width=4.5cm]{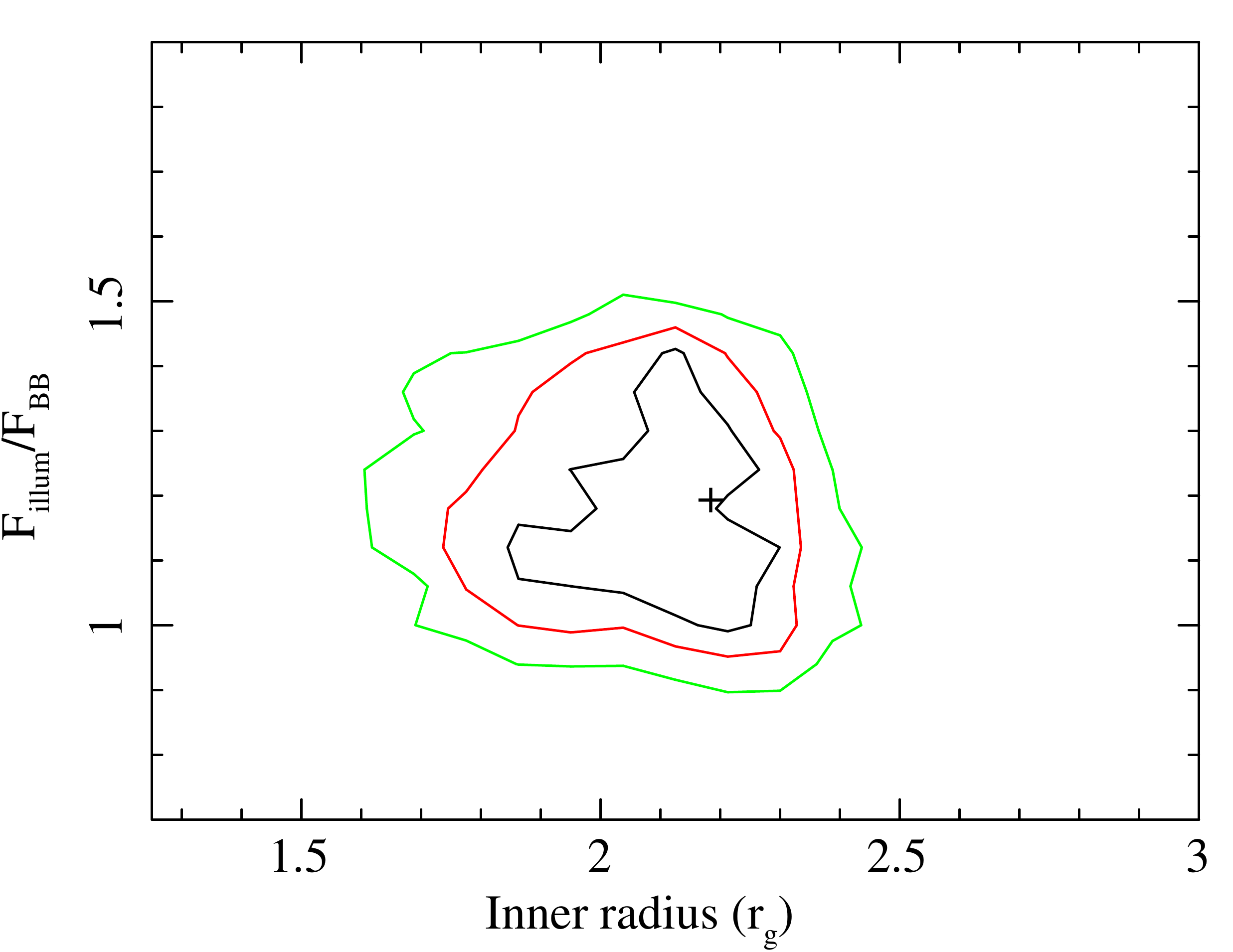}}
\hspace*{0.35cm}
\rotatebox{270}{\includegraphics[width=4.5cm]{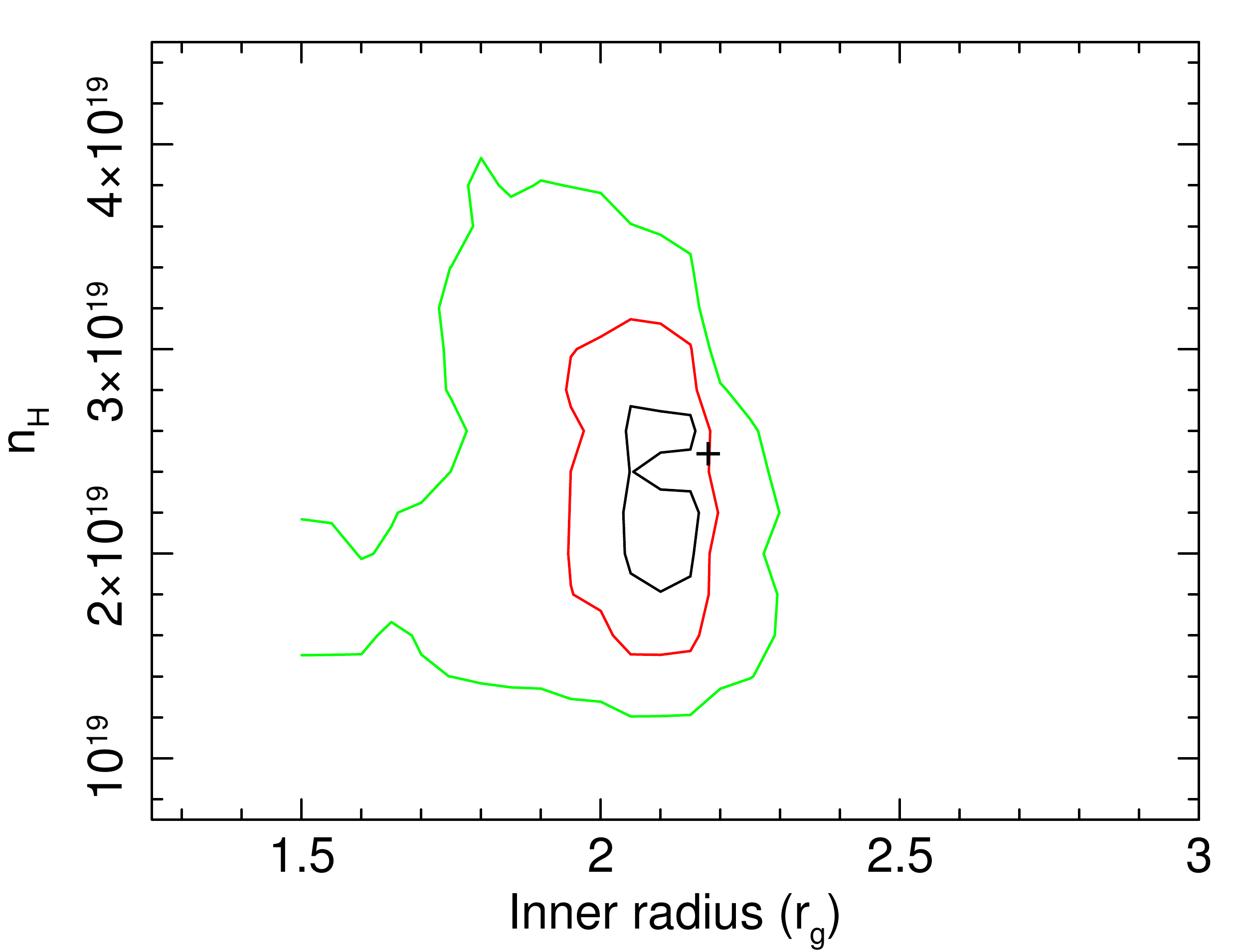}}
\caption{Contour plots investigating the effect of the inclination (top-left), inner (top-centre) and outer (top-right) emissivity indices, \pin/\xis\ cross-normalisation constant (bottom-left),  ratio of illuminating to black body flux (bottom-centre) and disk-surface hydrogen density (bottom-right)  on the inner radius for Model~5a. The latter two being a proxy to the common disk ionisation parameter.  The 68, 90 and 95 per cent confidence range for two parameters of interest are shown in black, red and green respectively.  The cross marks the global minima. It is clear that the inner extent of the accretion disk is robust to these parameters and is well constrained to $\sim1.9-2.3$~\rg\ at the 90~per~cent level of confidence.  }
\label{fig7}
\end{figure*}

A question that generally arises when one is dealing with reflection models is whether the constraints on the various parameters of interest (i.e. inner radius/spin, disk inclination  and emissivity profile, etc)  are driven by anything other than the reflection features. For example; could it be that extreme blurring, which would suggest a maximally rotating black hole, is artificially caused by the model trying to ``smooth" the soft part of the reflection spectrum to mimic a disk component?  To test this we ignored the \xis\ data below 3\kev\ where the disk emission dominates (see Fig.~\ref{fig2}) and removed the \pin\ data altogether. What remained is essentially the broad feature which we are associating with iron fluorescence emission. We refit Model~4a after freezing the  disk temperature and its normalization as well the neutral hydrogen column density and powerlaw index, as these cannot be constrained from the line profile alone. We refer to this as Model~4b in Table~2, and show this fit in Fig.~\ref{fig6}. It is clear that all parameters obtained from this narrow energy range are consistent, within errors, to that obtained using the full spectra. Allowing $\Gamma$ to vary does not change this conclusion.

Until this point, we have made use of \reflionx\ which is designed to reproduce reflection spectra from the accretion disks around AGN. However, the disks around stellar mass black holes are significantly hotter, resulting in subtle differences in the radiation processes, including reprocessing.  The higher disk temperatures around stellar mass black holes means that  Compton broadening is of greater importance and should therefore be included at the correct level.  With this in mind, \citet{refbhb}  developed a modified version of the previous grid, \refbhb,  in which the atmosphere of the accretion disk is is illuminated not only by the hard, powerlaw-like corona, but also by a further blackbody radiation intrinsic to the disk. This model self-consistently accounts for the disk, relativistic line and reflection continuum present in the phenomenological models described in the previous section.

The parameters of the model are the number density of hydrogen in the
illuminated atmosphere, $n_{H}$, the temperature of the accretion disk, $kT_{disk}$ the index of the (assumed powerlaw) continuum and the ratio of the total flux illuminating the disc to the total
blackbody flux emitted by the disc, $F_{illum}/F_{BB}$. The disc reflection spectra is again convolved with \kdblur\ to account for relativistic effects. This model (Model~5a; Table~2 and shown in Figure~5) provides the best fit yet to the data with $\chisq/\nu=344.9/322$ despite being more constrained --- compared to purely phenomenological models --- by virtue of being physically self-consistent. The inner accretion disk radius of $r_{in} = 2.2^{+0.2}_{-0.4}$\rg\  is  consistent with the values found from the \reflionx\ model together with a separate disc component. Assuming that this radius is the same as the radius of the innermost stable circular orbit, we can constrain the spin parameter \citep{Bardeenetal1972} to be $a=0.915^{+0.05}_{-0.03}$.

\begin{figure}
{\hspace*{-0.5cm}
 \rotatebox{0}{
{\includegraphics[width=8.5cm]{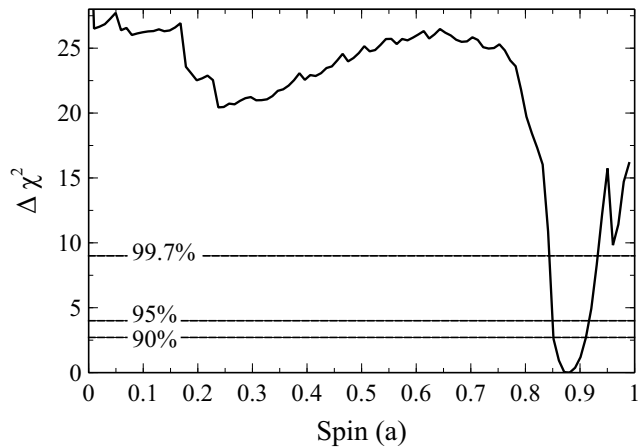}  
}}}
\caption{Goodness-of-fit versus spin for Model~5b. It is clear that the black hole in \maxi\ is rapidly rotating.  A maximally rotating Kerr or a static Schwarzschild black hole are both rejected at  greater than the 3$\sigma$ level of confidence.}
\label{fig8} 
\end{figure}

{  We again investigate the dependence of the inner radius \rin\ on a number of key parameters including the  inclination angle, the inner and outer emissivity indices, the cross normalisation between the \xis\ and \pin\ data, the ratio of the illuminating powerlaw to the blackbody flux  and the hydrogen number density at the disk surface.  Figure 7 shows that the  inner radius found here of  of $r_{in} = 2.2^{+0.2}_{-0.4}$\rg\ is extremely robust to changes in all the aforementioned parameters. }However, in order to make a formal constraint on the spin, we replace \kdblur\ with the  sophisticated variable-spin relativistic smearing model \relconv\ \citep{relconv}.  Figure~8 shows that  the spin of the black hole in \maxi\ is  well constrained to be $a=0.88\pm0.03$ at the 90~per~cent confidence range. In the following, section we will discuss, amongst other things,  the current strength and limitation of the various models used throughout this work and highlight some of the factors contributing to the tight constraint on the spin parameter of \maxi.

\section{discussion}
\label{discussion}

The fractional contribution of the disk component to the total 2--20\kev\ flux ($\sim25~\%$) places the current observation in the \textit{hard/intermediate} state as defined by  \citet{Bellonibook2010} and in an intermediate state between the \textit{low/hard} and \textit{high/soft} state definition of \citet{mcclintock06} and. However, we can see from  figures 5 and 6 that the \emph{observed} continuum is not dominated by the powerlaw-like component  expected to originate from the corona, but rather it is mostly reflection dominated. Similar ``reflection-dominated" spectra are seen in a number of narrow line Seyfert 1s and  quasars during their \textit{low state} (e.g. 1H0419-577, \citet{Fabian04192005}; NGC4051, \citet{Ponti06}; PG1543+489, \citet{Vignali2008}; Mrk 335, \citet{Grupe08}; PG1535+547,  \citet{Ballo2008}; PG2112+059, \citet{Schartel2010}). In this scenario, the majority of the X-ray reprocessing (reflection) occurs in the inner region of the accretion disk where  strong gravitational light bending is expected to occur \citep{MartocchiaMatt1996, Miniu04}.  Such behaviour is expected as a result of  strong light bending, where the reflected flux is enhanced over the inner regions as a result of gravitational focusing of the X-ray continuum down towards the black hole and onto the disk. The decrease in the number of X-rays that can escape as part of the continuum thus causes the source to appear reflection dominated.  

\citet{wilkins2011} showed  that a possible consequence  of strong gravitation effects is an increase in the emissivity profile of the disk. Classically, the emissivity profile is expected to be flat in the region directly  below the source, while tending to $r^{-3}$  when $r>>h$, where the flux received by the disc from the source falls off as the inverse square of the distance with a further factor of $1/r$ arising from the cosine of the angle projecting the ray normal to the disc plane. However, as briefly mentioned above and detailed in \citet{wilkins2011}, strong gravity can potentially  act to focus more of the direct continuum into the inner parts of the disk as well as increase the disk area being radiated -- the latter as a consequence of  gravitational warping. In this particular treatment, these factors  causes for a substantial steepening of the emissivity profile in the inner regions. 

This is indeed a possible explanation for what is observed in \maxi, where within a radius of $\lesssim 4$\rg\ the emissivity index is consistently $>6$ (see Table~2) and beyond it goes closer to the classical value of 3.  Both the reflection-dominated spectrum and the high emissivity profile seen here suggest that the primary X-ray continuum is located within a few gravitational radii of the black hole. Looking at Figure~1, it is indeed possible that the corona briefly ``collapsed" down close to the black hole  causing the decrease in the hard X-ray flux seen during this \suzaku\ observation. If this is the case, state transitions (at least that between the \textit{high/soft} and \textit{low/hard})  should be seen to be much more tightly  associated with changes in the corona as opposed to physical changes in the accretion disk.

\begin{figure}
{\hspace*{-0.cm}
 \rotatebox{0}{
{\includegraphics[width=8.5cm]{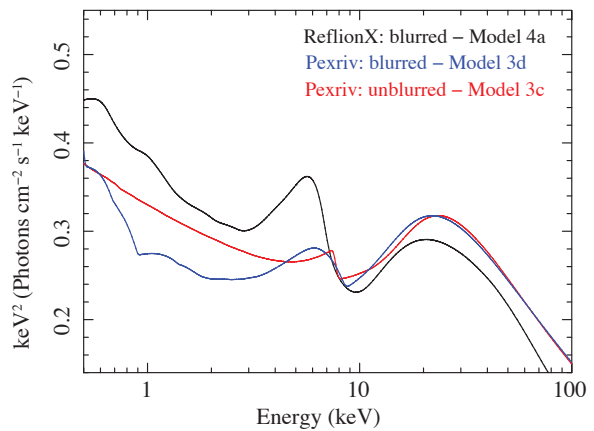}  
}}}
\caption{Difference between the blurred \reflionx\ component in Model~4a (black) to the \pexriv\ components in Models 3d (blue) and 3c (red). The unblurred \pexriv\ (red) is as expected from a ``cold" ($\xi=70\ergcmps$) accretion disk being illuminated by a $\Gamma = 2.18$ powerlaw without being relativistic convolved. Compare this to the component expected from  a  moderately ionised ($\xi=5200\ergcmps$; blue) disk at 3.3\rg\ illuminated by a $\Gamma = 2.09$ powerlaw and the \reflionx\ model having both the absorption edge as well as the iron emission line (black).  }
\label{fig9} 
\end{figure}

In this work we have investigated a variety of possible models striving towards a physically motivated interpretation for the observed spectrum.  It is worth stressing that  equally good fits -- and at times (albeit not in this work) statistically better fits -- can be obtained with  a purely phenomenological combination of components severely lacking in physical consistency.  A case in point is Model~3c where the presence of the -- clearly broad -- iron line requires the emission to occur from deep within 6\rg, yet  the direct X-ray and reflection continuum appear somehow exempt from the effect of strong gravity. When we account for relativistic effects,  the ionization parameter of the disk increases by nearly two orders of magnitude from $\xi \approx 70$ to $\approx5000\ergcmps$. This change is accompanied by a hardening of the photon index $\Gamma$ possibly due to the powerlaw  trying  to compensate for the stronger soft emission and the weaker Compton hump of the blurred ionised reflector  in comparison to the unblurred, cold reflector (see Fig.~\ref{fig9}). 

A further point to note is that the  apparent  decrease in the equivalent width of the relativistic line profile (from 270\ev to 180\ev\  from  model 3c to model 3d) is clearly a consequence of the,  physically inconsistent, lack of gravitational blurring. From Fig.~\ref{fig9} we can see that the \textit{absorption edge} in \pexriv\ becomes much more smooth and symmetric after the component is convolved with \kdblur\ and the ionization increases. Due to the decoupling between the \laor\ line profile and \pexriv, this smooth edge can act somewhat like an emission line conspiring against the  \laor\ line component  and thus decreasing the equivalent width of latter. In order to couple the emission line strength with the absorption edge and all other reflection features, we replaced the phenomenological combination of \laor\ + \pexriv\ with the \reflionx\ reflection grid. The first thing to notice from Table~2 (model 4) is that this recovered the high ionization value from Model~3d but now in a self-consistent manner. Nonetheless, the strong requirement for an emission line with an equivalent width  of at least  $\approx180$\ev\ (Model~3d) is  highly  indicative  of a broad line, adhering to the strong criterion of  \citet{reislhs} arguing against an accretion disk truncated far from the ISCO.

The precise value of the neutral hydrogen column density towards \maxi\ is not known, however, based on Swift/XRT data which extends to much lower energies as compared to the current \suzaku\ observation,  \citet{atel3613} showed that a likely value is  $(2\pm0.4)\times10^{21} \pcmsq$. A  similar value is found here using our best model. Figure~9 highlights how differing models impact the energies below $\sim1\kev$ where we would begin to see the curvatures expected from such a low column density. The fact that our data cuts off at 1.2\kev\ means that, in order to break the  degeneracies between the -- essentially disk -- continuum (see Fig.~2) and neutral column density, one must use a self consistent model for the full spectra. Such is the case for \refbhb.

It is well known that in the hot inner regions of an accretion disk the observed disk spectrum suffers from the effects of electron scattering which results in an observed (colour) temperature, $T_{col}$,  which is higher than the effective blackbody temperature, $T_{bb}$,  by approximately a factor of $f_{col} = T_{col}/T_{bb}$ \citep{RossFabianMineshige1992}. This ``colour correction factor" has been shown to have  a value of $1.7\pm0.2$ \citep{ShimuraTakahara1995}  for a wide range in luminosity, as long as the disk effective temperature remains below $\sim1$\kev\ \citep{DavisBlaes2005colour}, as is the case here. Above this temperature, disk self ionisation can lead  to an increase in  $f_{col} $ however it is found to be consistently below $\sim 3$ \citep{merlonifabianross00}.  The effective temperature in the \refbhb\ (Model~5) of  $kT_5 = 0.27^{+0.02}_{-0.01}$\kev\ is precisely as expected from the value of the  colour temperature from the \diskbb\  component in Model~4 ($kT_4/kT_5 = 1.7\pm0.1$) .

Lastly, although we cannot directly compare $\xi$ between \reflionx\ and \refbhb, we can estimate this value based on the various parameters output in Table~{2}. Since the flux of the blackbody in the model is related to its temperature by the  $F_{bb} \propto  T^4$ relation, we find $F_{bb}= (5.5^{+1.8}_{-0.8}\times10^{21}\ergpcmsqps$.  Combining this $F_{Illum}/F_{BB}=1.2\pm0.2$  gives an illuminating flux of $(6.5^{+2.4}_{-1.4})\times10^{21}\ergpcmsqps$. The ionization parameter, which is defined as $\xi = 4\pi F/n ~\ergcmps$, is  then found by using 
$n = n_H = 2.5^{+0.7}_{-1.0} \times 10^{19}{ \rm H~cm^{-3}}$. In this manner, we find $\xi = 3300\pm1500 ~\ergcmps$, in perfect agreement with the values found for both \reflionx\ in both Models~4a and 4b and the blurred \pexriv\ in Model~3d.

All current methods of measuring black hole spin rely on both the assumption that the accretion disk extends to the innermost stable circular orbit, as well as that emission within this radius is negligible.  The latter is indeed valid for the standard model of black hole accretion \citep{shakurasunyaev73}, where within the ISCO -- in the region often referred to as the plunging region -- there is no angular momentum transport, and the region cannot support an X-ray corona to irradiate the material that is ballistically plunging onto the black hole. However, it was shown by \citet{Krolik1999} and independently by \citet{Gammie1999} that when magnetic fields are considered, the B-fields within the ISCO may be amplified to a point where the magnetic energy density can be comparable to the rest mass energy of the accreting material and may lead to the creation of an active inner X-ray corona. \citet{Shafee2008} investigated the
effect of magnetic torque within the plunging region around a non-rotating black hole and concluded that
``...magnetic coupling across the ISCO is relatively unimportant for geometrically-thin discs". 

A further
study specifically aimed at addressing the robustness of the iron line/reflection fitting technique in diagnosing
black hole spin was presented by \citet{reynoldsfabian08}. The authors used a high resolution
three-dimensional magnetohydrodynamic simulation of a geometrically-thin accretion disc to show that
the density of the plunging material drops precipitously over a very small radius within the ISCO. This
sudden drop in density results in the material being highly photoionised and suppresses
any significant iron line emission as well as all other reflection features from within the ISCO.
The study by \citet{reynoldsfabian08} concluded that for a non-rotating black hole where the ISCO is at 6\rg, the “reflection edge” -- defined by the authors as the innermost radius from which significant reflection
emission is seen -- is at approximately 5.8\rg. Furthermore, the discrepancy between the true ISCO and
the inferred radius diminishes as one considers more rapidly rotating black holes as is the case in \maxi\ (see Fig.~8).  From an observational perspective, it is also worth noting that strong support for the presence of an ``inner edge"  in the accretion disk surrounding black holes is provided by decades of empirical evidence as  shown in \citet{steiner2010}.

The culmination of this work is that the recently discovered system, \maxi, is indeed a stellar mass black hole binary, having a central black hole rotating with a spin parameter of $a=0.88\pm0.03$ (90\% confidence). This strong constraint is a result of being able to successfully, and most importantly self-consistently, model the reflection features clearly present in the \suzaku\ spectra.

\section{Acknowledgements}
RCR thanks the Michigan Society of Fellows and NASA. A further thank you goes to the \suzaku\ team, which scheduled this TOO observation, and to the MAXI team for providing prompt notice of this new source to the astronomical community at large.  RCR is supported by NASA through the Einstein Fellowship Program, grant number PF1-120087 and is a member of the Michigan Society of Fellows. ACF thanks the Royal Society. This work was greatly expedited thanks to the help of Jeremy Sanders in optimising the various convolution models.

\bibliographystyle{/Users/rdosreis/papers/maxij1836-194/paper_afteref/mnras}

\bibliography{/Users/rdosreis/papers/bibtex.bib}
\end{document}